\documentclass[aps,pra,amsfonts,amssymb,amsmath,showpacs,
floatfix,nofootinbib,groupedaddress,superscriptaddress,citesort]{revtex4}
\usepackage{subfigure}
\usepackage{mathrsfs}
\usepackage{longtable}
\usepackage{amsfonts}
\usepackage{amstext}
\usepackage{amsmath}
\usepackage{amssymb}
\usepackage{amsthm}
\usepackage[usenames]{xcolor}
\usepackage[dvips]{graphicx}
\def\qed{\leavevmode\unskip\penalty9999 \hbox{}\nobreak\hfill
     \quad\hbox{\leavevmode  \hbox to.77778em{%
              \hfil\vrule   \vbox to.675em%
               {\hrule width.6em\vfil\hrule}\vrule\hfil}}
     \par\vskip3pt}

\newtheorem{lemma}{Lemma}
\newtheorem{thm}[lemma]{Theorem}
\newtheorem{*thm*}[lemma]{Theorem}
\newtheorem{prop}[lemma]{Proposition}

\begin{document}
\title{A note on the Bloch representation of absolutely maximally entangled states}

\author{Bo Li}
\affiliation{School of Mathematics and Computer science, Shangrao Normal University,
 Shangrao 334001, China}\affiliation{Max-Planck-Institute
for Mathematics in the Sciences, 04103 Leipzig, Germany}
\author{Shuhan Jiang}
\affiliation{Max-Planck-Institute
for Mathematics in the Sciences, 04103 Leipzig, Germany}
\affiliation{School of Physics, Nankai University, Tianjin 300071, China}
\author{Shao-Ming Fei}
 \affiliation{Max-Planck-Institute
for Mathematics in the Sciences, 04103 Leipzig, Germany}\affiliation{School of Mathematical Sciences, Capital Normal University, Beijing 100048, China}
\author{Xianqing Li-Jost}
\affiliation{Max-Planck-Institute
for Mathematics in the Sciences, 04103 Leipzig, Germany}

\begin{abstract}
The absolutely maximally entangled (AME) states play key roles in quantum information processing.
We provide an explicit expression of the generalized Bloch representation of AME states for general dimension $d$ of individual subsystems and arbitrary number of partite $n$.
Based on this analytic formula, we prove that the trace of the squared support for any given weight is given by the so-called hyper-geometric function and is
irrelevant with the choices of the subsystems. The optimal point for the existence of AME states is obtained.
\end{abstract}
\maketitle

Multipartite entanglement is not only an interesting phenomenon in quantum physics, but also a key
resource in quantum information theory, which can allow for novel quantum advantages in quantum information processing. One of the most striking phenomena in multipartite entangled systems is that
although one knows the completed knowledge of the whole system, one can not say all the knowledge of its subsystems \cite{Nielsen&Chuang}.
In particular, for a maximally or genuinely entangled multipartite pure state, its reduced subsystems may be maximally mixed ones.

The family of absolutely maximally entangled (AME) states is the class of $n$-partite pure states such that all of the reduced $\lfloor n/2\rfloor$-partite states are maximally mixed \cite{Gisin1998,Higuchi2000}.
AME states play an important role in quantum information processing like quantum teleportation \cite{Helwig2012,Helwig2013,Helwig2013_graph} and
quantum error correction \cite{Scott2004,raissi,Nebe2006,Grassl2015,Calderbank1998,Zha2011}.
AME states have also deep connections with apparently irrelevant areas of mathematics such as combinatorial designs \cite{Goyeneche2015} and holography \cite{Latorre2015,Pastawski}.
Furthermore, AME states are the special cases of $k$-uniform states for $k=\lfloor n/2\rfloor$ \cite{Arnoud2013,Goyeneche2014}, while the $k$-uniform states
also play a central role in quantum error correction \cite{Feng2015}.

A well known related open question is to determine the existence of AME states for given dimension and number of partites \cite{Bernal2017}.
For multiqubit systems (dimension $d=2$), the two-qubit Bell state, the three-qubit Greenberger-Horne-Zeilinger state are the AME states since all the one-qubit
reduced density matrices are maximally mixed. However, the four-qubit AME states are shown to be not existing \cite{Higuchi2000,Gour2010}. The five and six-qubit AME states
are constructed explicitly in \cite{Scott2004,Borras2007,Facchi2010}.
While it has been shown that AME states do not exist
for more than eight-qubit systems \cite{Rains1998,Rains1999},
the existence of AME states for seven-qubit systems had been a longstanding open problem.
Very recently, Huber \emph{et.al} proved that seven-qubit AME states do not exist \cite{Huber2017},
which completely solves the existence problem of AME states for multi-qubit case.

For the case that the dimension of each individual subsystem is great than two ($d>2$), it has been shown that the AME states exist for any multipartite systems for suitably chosen $d$. These states can be constructed from graph states, and used for various quantum information tasks \cite{Helwig2012,Helwig2013,Helwig2013_graph,Hein2006,Steinhoff,Chen2016}.
Partial results have been also obtained for particular cases such as even number of partite \cite{Goyeneche2014} and with minimal support \cite{Goyeneche2015}.
For general $d$ and $n$-partite systems, a necessary condition for the existence of AME states
has been presented in \cite{Hein2006,Scott2004},
\begin{eqnarray}\label{nineqorigin}
n\leq\left\{
\begin{aligned}
& 2(d^2-1) & ~~~~~~& n~ even,\\
& 2d(d+1)-1 & ~~~~~~ &n~ odd.
\end{aligned}
\right.
\end{eqnarray}
For systems such that $n$ and $d$ do not satisfy the above inequality,
there will be no AME states.

In this note, from a detailed analysis on the generalized Bloch representation of AME states, we provide explicit formulae for the operator $P_{\lfloor n/2\rfloor+i}^2$ appeared
in the generalized Bloch representation of AME states for arbitrarily $n$ partite system with each dimension $d$. We derive two striking things,
one is that the critical point for the existence of AME states is always $i=2$ by taking into account the positivity of
$tr(P_{\lfloor n/2\rfloor+i}^2)$, the another is that the eigenvalues of $P_{\lfloor n/2\rfloor+i}$ should always be positive.
Our method generalize the one used in \cite{Huber2017} and can be used to estimate the existence of AME states.



An AME state \cite{Helwig2013,Helwig2013_graph} is a pure state of $n$-partite, $P=\{1,\ldots,n\}$, with each dimension $d$. Let $H_i$, $H_i\cong \mathcal{C}^d$, $i=1,...,n$, denote $d$-dimensional vector spaces.
Consider a pure state $|\Phi\rangle\in  H_1\otimes\ldots\otimes H_n$. Under a bipartite partition
$A$ and $B$, $A\cup B=P$, the state $|\Phi\rangle$ can be written in the following Schmidt form,
\begin{eqnarray*}
|\Phi\rangle=\sum_{i=1}^{d^k}\sqrt{d^{-k}}|\phi_i\rangle_A \otimes|\psi_i\rangle_{B_1}\otimes|\psi_i\rangle_{B_2}\otimes\ldots\otimes|\psi_i\rangle_{B_k},
\end{eqnarray*}
where B is a $k$-partite system and A is an $(n-k)$-partite system such that $k\leq n-k$,
$\langle\phi_i|\phi_{i'}\rangle_A=\delta_{ii'}$, and $|\psi_i\rangle_{B_k}$ is the local orthogonal base of the subsystem $B_k$. The $(n-k)$-partite reduced state $\rho_A$ is given by
\begin{eqnarray}\label{rhoa}
\rho_A=\sum_{i=1}^{d^k}d^{-k}|\phi_i\rangle_A\langle\phi_i|,
\end{eqnarray}
which has $d^k$ nonzero eigenvalues $\lambda=d^{-k}$. By definition, the reduced state
$\rho_B$ needs to be maximally mixed,
\begin{eqnarray*}
\rho_B=\sum_{i=1}^{d^k}d^{-k}|\phi_i\rangle_{B_1}\langle\phi_i|
\otimes\ldots\otimes|\phi_i\rangle_{B_k}\langle\phi_i|=d^{-k}I_{d^{k}}
\end{eqnarray*}
for $k=\lfloor n/2\rfloor$.

It is obvious that if the  Schmidt form of $|\Phi\rangle$ holds true for $k=\lfloor n/2\rfloor$, then it is also true for all $k\leq\lfloor n/2\rfloor$.
From Eq.(\ref{rhoa}), we can obtain the following projector property
\begin{eqnarray}\label{projectorpro}
\rho_A^2=d^{-k}\rho_A,
\end{eqnarray}
where $\rho_A$ is any $n-k$ reduced state of $|\Phi\rangle$, $0<k\leq\lfloor n/2\rfloor\leq n-k$.

By Bloch representation any $n$-qudit state can be written as
\begin{eqnarray}\label{Blochrepresentation}
\rho=d^{-n}\sum_{\alpha_1\ldots\alpha_n}r_{\alpha_1\ldots\alpha_n}\lambda_{\alpha_1}\otimes\ldots\otimes\lambda_{\alpha_n},
\end{eqnarray}
where $\alpha_i=0,1,\ldots,d^2-1$, $i=1,...,n$, $r_{\alpha_1\ldots\alpha_n}$ are real coefficients, and $\lambda_{\alpha_i}$ are the traceless Hermitian generators of $SU(d)$.
For simplicity, we group the terms in (\ref{Blochrepresentation}) according to their weights, the number of nontrivial $SU(d)$ generators.
Let $P_j$ denote the summation of all the terms containing $j$ number of nontrivial $SU(d)$ generators
(the remaining part are all identities), see \cite{Huber2017}. We can rewrite the state $\rho$ as
\begin{eqnarray}\label{weightj}
\rho=d^{-n}(I+\sum_{j=1}^{n}P_j),
\end{eqnarray}
where, and also in the following, $I$ denotes the corresponding identity operator.

If $\rho$ is an AME state, then all $\lfloor n/2\rfloor$-partite reduced states of $\rho$ are maximally mixed. Namely, in (\ref{weightj}) all $P_j=0$ for $j\leq \lfloor n/2\rfloor$. Then the $(\lfloor n/2\rfloor+1)$-partite reduced state of $\rho$ is given by
\begin{eqnarray}\label{projectorproper133}
\rho_{(\lfloor n/2\rfloor+1)}=d^{-(\lfloor n/2\rfloor+1)}(I+P_{(\lfloor n/2\rfloor+1)}).
\end{eqnarray}
On the other hand, from (\ref{projectorpro}) we have
\begin{eqnarray}\label{projectorproper1}
\rho_{(\lfloor n/2\rfloor+1)}^2=d^{-(n-(\lfloor n/2\rfloor+1))}\rho_{(\lfloor n/2\rfloor+1)}.
\end{eqnarray}

Substituting (\ref{projectorproper133}) into (\ref{projectorproper1}) and tracing over the whole system, we obtain that for every $(\lfloor n/2\rfloor+1)$-partite
subsystem, the $tr(P_{\lfloor n/2\rfloor+1}^2)$ are all equal and is given by
\begin{eqnarray*}
tr(P_{\lfloor n/2\rfloor+1}^2)=d^{\lfloor n/2\rfloor+1}(\frac{d^{2(\lfloor n/2\rfloor+1)}}{d^n} -1),
\end{eqnarray*}
which means that the term $P_{\lfloor n/2\rfloor+1}$ exists in any $(\lfloor n/2\rfloor+1)$-partite reduced density matrices, and moreover, $tr(P_{\lfloor n/2\rfloor+1}^2)$ is irrelevant with the
choices of $\lfloor n/2\rfloor+1$ subsystems.


Now let us consider $tr(P_{\lfloor n/2\rfloor+2}^2)$ in the $(\lfloor n/2\rfloor+2)$-partite reduced state $\rho_{(\lfloor n/2\rfloor+2)}$. $\rho_{(\lfloor n/2\rfloor+2)}$ satisfies the property (\ref{projectorpro}) and has the Bloch representation with all $P_j=0$ for $j\leq \lfloor n/2\rfloor$.
The terms $P_{\lfloor n/2\rfloor+1}$ existed in $\rho_{(\lfloor n/2\rfloor+1)}$ all appear in $\rho_{(\lfloor n/2\rfloor+2)}$. By tracing the equation
$\rho_{(\lfloor n/2\rfloor+2)}^2=d^{-(n-(\lfloor n/2\rfloor+2))}\rho_{(\lfloor n/2\rfloor+2)}$, while
noticing that $tr(( P_{\lfloor n/2\rfloor+2})(\sum P_{\lfloor n/2\rfloor+1}\otimes I))=0$, we can prove that the term $P_{\lfloor n/2\rfloor+2}$ exists in any $(\lfloor n/2\rfloor+2)$-partite reduced density matrices and $tr(P_{\lfloor n/2\rfloor+2}^2)$ is irrelevant with the
choices of $\lfloor n/2\rfloor+1$ subsystems.

By mathematical reduction, we can prove for any $\lfloor n/2\rfloor+i$
reduced subsystem, $tr(P_{\lfloor n/2\rfloor+i}^2)$ are all equal and irrelevant with the
choices of $\lfloor n/2\rfloor+i$ subsystems. Moreover, the number of terms like $P_{\lfloor n/2\rfloor+1}\otimes I$ in the summations $\sum P_{\lfloor n/2\rfloor+1}\otimes I$ in $\rho_{(\lfloor n/2\rfloor+i)}$ is $\binom{\lfloor n/2 \rfloor+i}{\lfloor n/2 \rfloor+1}$.
We have the following Proposition.
\begin{prop}
Let $|\Phi\rangle\in  H_1\otimes\ldots\otimes H_n$ be an AME state. The density matrix $|\Phi\rangle\langle|\Phi|$
is given by
\begin{eqnarray}
\rho  & = &d^{-n}(I+\sum_{j_1\ldots j_{\lfloor n/2\rfloor+1}}P_{\lfloor n/2\rfloor+1}^{(j_1\ldots j_{\lfloor n/2\rfloor+1})}\otimes I^{\overline{(j_1\ldots j_{\lfloor n/2\rfloor+1})}}\nonumber\\
 & &+ \sum_{j_1\ldots j_{\lfloor n/2\rfloor+2}}P_{\lfloor n/2\rfloor+2}^{(j_1\ldots j_{\lfloor n/2\rfloor+2})}\otimes I^{\overline{(j_1\ldots j_{\lfloor n/2\rfloor+2})}})\nonumber\\
  &  &+\ldots +P_n), \label{kuniformstte}
\end{eqnarray}
where, e.g., $P_{\lfloor n/2\rfloor+1}^{(j_1\ldots j_{\lfloor n/2\rfloor+1})}\otimes I^{\overline{(j_1\ldots j_{\lfloor n/2\rfloor+1})}}$ denotes terms with $\lfloor n/2\rfloor+1$ nontrivial generators acting on the $j_1\ldots j_{\lfloor n/2\rfloor+1}$ subsystems, and $I^{\overline{(j_1\ldots j_{\lfloor n/2\rfloor+1})}}$ represent the corresponding identities $I$ on the remaining subsystems. The value of $tr((P_{\lfloor n/2\rfloor+1}^{(j_1\ldots j_{\lfloor n/2\rfloor+1})})^2)$ are all equal and independent on
choices of $j_1\ldots j_{\lfloor n/2\rfloor+1}$. There are $C_n^{\lfloor n/2\rfloor+1}$ terms
like $P_{\lfloor n/2\rfloor+1}^{(j_1\ldots j_{\lfloor n/2\rfloor+1})}\otimes I^{\overline{(j_1\ldots j_{\lfloor n/2\rfloor+1})}}$ in the summation $\sum_{j_1\ldots j_{\lfloor n/2\rfloor+1}}P_{\lfloor n/2\rfloor+1}^{(j_1\ldots j_{\lfloor n/2\rfloor+1})}\otimes I^{\overline{(j_1\ldots j_{\lfloor n/2\rfloor+1})}}$, and $C_n^{\lfloor n/2\rfloor+2}$ terms like
$P_{\lfloor n/2\rfloor+2}^{(j_1\ldots j_{\lfloor n/2\rfloor+2})}\otimes I^{\overline{(j_1\ldots j_{\lfloor n/2\rfloor+2})}})$ in the summation $\sum_{j_1\ldots j_{\lfloor n/2\rfloor+2}}P_{\lfloor n/2\rfloor+2}^{(j_1\ldots j_{\lfloor n/2\rfloor+2})}\otimes I^{\overline{(j_1\ldots j_{\lfloor n/2\rfloor+2})}})$ and so on.
\label{proposition}
\end{prop}
\par

In addition, taking into account that $tr(P_sP_t)=0$ for $s\neq t$, and
\begin{eqnarray*}
 &  &tr((\sum_{j_1\ldots j_{\lfloor n/2\rfloor+1}}P_{\lfloor n/2\rfloor+1}^{(j_1\ldots j_{\lfloor n/2\rfloor+1})}\otimes I^{\overline{(j_1\ldots j_{\lfloor n/2\rfloor+1})}})\cdot\nonumber\\
 & & (\sum_{j_1\ldots j_{\lfloor n/2\rfloor+1}}P_{\lfloor n/2\rfloor+1}^{(j_1\ldots j_{\lfloor n/2\rfloor+1})}\otimes I^{\overline{(j_1\ldots j_{\lfloor n/2\rfloor+1})}}))\nonumber\\
 & =& C_n^{n-(\lfloor n/2\rfloor+1)}\times(n-(\lfloor n/2\rfloor+1))!\nonumber\\
  &  &\times d^{n-(\lfloor n/2\rfloor+1)}\times trP_{\lfloor n/2\rfloor+1}^2,
\end{eqnarray*}
we have
\begin{eqnarray}
tr(\rho^2)& = &  d^{(-2n)}(d^n+ \binom{n}{\lfloor n/2 \rfloor+1} \times d^{n-(\lfloor n/2\rfloor+1)}  \nonumber\\
 & &\times tr(P_{\lfloor n/2\rfloor+1}^2)+  \binom{n}{\lfloor n/2 \rfloor+2} \times d^{n-(\lfloor n/2\rfloor+2)}\nonumber\\
  &  & \times tr(P_{\lfloor n/2\rfloor+2}^2)+\ldots+tr(P_{n}^2)), \label{tracesrho2}
\end{eqnarray}
which is equal to one, as for any pure state $\rho$, $tr(\rho^2) =1$.


Note that for $k\geq \lfloor n/2\rfloor+1$, the reduces states still
satisfy the Eq.(\ref{projectorpro}). In particular, let us consider the $\lfloor n/2\rfloor+i$
reduced states $\rho_{(\lfloor n/2\rfloor+i)}$  of $\rho$, where $i\geq 1$, $\lfloor n/2\rfloor+i<n$.
Since $\rho$ is an AME state, similar to Eq. (\ref{kuniformstte}), $\rho_{(\lfloor n/2\rfloor+i)}$  can be expressed as
 \begin{eqnarray}
   &  & \rho_{(\lfloor n/2\rfloor+i)}  =d^{-(\lfloor n/2\rfloor+i)}(I+  \nonumber\\
 & &   \sum_{j_1\ldots j_{\lfloor n/2\rfloor+1}}P_{\lfloor n/2\rfloor+1}^{(j_1\ldots j_{\lfloor n/2\rfloor+1})}\otimes I^{\overline{(j_1\ldots j_{\lfloor n/2\rfloor+1})}} \nonumber\\
 &  & +\sum_{j_1\ldots j_{\lfloor n/2\rfloor+2}}P_{\lfloor n/2\rfloor+2}^{(j_1\ldots j_{\lfloor n/2\rfloor+2})}\otimes I^{\overline{(j_1\ldots j_{\lfloor n/2\rfloor+2})}}\nonumber\\
  &  & +\ldots +P_{(\lfloor n/2\rfloor+i)}), \label{kuniformstteee}
\end{eqnarray}
It should be noted that the term $\sum_{j_1\ldots j_{\lfloor n/2\rfloor+1}}P_{\lfloor n/2\rfloor+1}^{(j_1\ldots j_{\lfloor n/2\rfloor+1})}\otimes I^{\overline{(j_1\ldots j_{\lfloor n/2\rfloor+1})}}$ in (\ref{kuniformstteee})  are $(\lfloor n/2\rfloor+i)$-partite states .
Combining Eq. (\ref{tracesrho2}) and (\ref{kuniformstteee}) we have
\begin{eqnarray}\label{uni2}
tr(\rho_{(\lfloor n/2\rfloor+i)}^2 )=d^{-2(\lfloor n/2\rfloor+i)}(d^{\lfloor n/2\rfloor+i}+u^2),
\end{eqnarray}
where
\begin{align}\label{u2}
  u^2 & =\binom{\lfloor n/2 \rfloor+i}{\lfloor n/2 \rfloor+1} d^{i-1} tr(P_{\lfloor n/2\rfloor+1}^2)\\\nonumber
  &+\binom{\lfloor n/2 \rfloor+i}{\lfloor n/2 \rfloor+2} d^{i-2} tr(P_{\lfloor n/2\rfloor+2}^2) \\\nonumber
  &+\cdots+ tr(P_{\lfloor n/2\rfloor+i}^2).\nonumber
\end{align}

Eq.(\ref{uni2}) actually gives rise to a set of linear equations,
\begin{equation}
 \left \{
 \arraycolsep=0pt
 \begin{array}{rrrrl}
 A_{11}x_1&&&=&T_1,\\
 A_{21}x_1&+A_{22}x_2&&=&T_2,\\
 \cdots&\cdots&\cdots&&\\
  A_{i1}x_1&+A_{i2}x_2&+\cdots&+A_{ii}x_i=&T_i,\\
  \end{array}
 \right.
 \end{equation}
 where $x_j=tr(P_{\lfloor n/2 \rfloor+j}^2)$, $T_l=d^{-(n-(\lfloor n/2 \rfloor+l))}-d^{-(\lfloor n/2 \rfloor+l)}$, $j,l=1,\ldots,i$, and
 \begin{equation}
A_{lj}=\left\{
\begin{array}{ll}
d^{-2\lfloor n/2 \rfloor-l-j}\binom{\lfloor n/2 \rfloor+l}{\lfloor n/2 \rfloor+j},~~~~&1\leq j\leq l\leq i,\\
0.&1\leq l<j\leq i.
\end{array}
\right.
\end{equation}

Let $A$ denote the coefficient matrix with elements $A_{lj}$. The inverse matrix $A^{-1}$ can be obtained by substituting $d$ with $(-d)^{-1}$ for nonzero $A_{lj}$, and keeping the rest elements zero.
For $l<j$, one easily verifies that $(AA^{-1})_{lj}=0$. For $l\geq j$, we have
 \begin{eqnarray}
\begin{split}
(AA^{-1})_{lj}&=\sum_{k=1}^{i} A_{lk}A_{kj}^{-1}\\
&=d^{j-l}\sum_{k=j}^{l}(-1)^{k+j}\binom{\lfloor n/2 \rfloor+l}{\lfloor n/2 \rfloor+k}\binom{\lfloor n/2 \rfloor+k}{\lfloor n/2 \rfloor+j}\\
&=d^{j-l}\binom{\lfloor n/2 \rfloor+l}{\lfloor n/2 \rfloor+j}\sum_{k=j}^{l}(-1)^{k-j}\binom{l-j}{k-j}\\
&=d^{j-l}\binom{\lfloor n/2 \rfloor+l}{\lfloor n/2 \rfloor+j}(1-1)^{l-j}\\
&=\delta_{lj}.
\end{split}
\end{eqnarray}
Thus we have the following theorem

\begin{thm}\label{tracesquares}
Let $|\Phi\rangle$ be an AME state given in the form of (\ref{kuniformstte}). We have
\begin{eqnarray}
\begin{split}
tr(P_{\lfloor n/2 \rfloor+i}^2)&=\sum_{j=1}^{i}(A^{-1})_{ij}T_j \\
&=\frac{(-1)^i d^{i+\lfloor n/2 \rfloor}\binom{i+\lfloor n/2 \rfloor}{1+\lfloor n/2 \rfloor}\left( 1+\lfloor n/2 \rfloor-d^{2(1+\lfloor n/2 \rfloor)-n}(i+\lfloor n/2 \rfloor){}_{2}F_{1}(1,1-i;2+\lfloor n/2 \rfloor;d^2)\right)}{i+\lfloor n/2 \rfloor}, \label{p1}
\end{split}
\end{eqnarray}
where ${}_{2}F_{1}(a,b;c;z)$ is the so-called ordinary hyper-geometric function \cite{Hypergeometric}.
\end{thm}
\par
Theorem \ref{tracesquares} provide us an explicit formula to characterize the Bloch representation of AME states. For each $d$ and $n$,
$tr(P_{\lfloor n/2 \rfloor+i}^2)$ can be calculated easily and can be used to estimate the existence of AME states. For $i=2$, we have
\begin{eqnarray}
tr(P_{\lfloor n/2 \rfloor+2}^2)=
\begin{cases}
\frac{1}{2} (-1 + d) d^{(3 + n)/2} (-1 + 2 d + 2 d^2 - n) , &n~is~odd; \cr
\frac{1}{2} (-1 + d^2) d^{2 + n/2} (-2 + 2 d^2 - n), &n~is~even.\label{p2}
\end{cases}
\end{eqnarray}
The positivity of (\ref{p2}) yields the bound (\ref{nineqorigin}) given by Scott \cite{Scott2004}.  If for given $n$, $d$ and $i$,
$tr(P_{\lfloor n/2 \rfloor+i}^2)$ is negative, then we can rule out the existence of AME states in this case.
For $d=2$, by calculating $tr(P_{\lfloor n/2 \rfloor+i}^2)$ for different $n$ and $i$,
an interesting thing shows up that the negative value appears first for $i=2$, see Table~\ref{t1}.
\begin{table}[htbp]
	\caption{$tr(P_{\lfloor n/2 \rfloor+i}^2)$ for $n$-qubit AME states, $n=2,...,13$.}\label{t1}
	\vspace{0.5pt}
	\begin{center}
		\begin{tabular}{c c c c c c c c c c}
			\hline \hline
			$ $&$i=1$&$i=2$&$i=3$&$i=4$&$i=5$&$i=6$&$i=7$ \\
			$n=2$&$12$&$ $&$ $&$ $&$ $&$ $&$ $ \\
			$n=3$&$4$&$32$&$ $&$ $&$ $&$ $&$ $ \\
			$n=4$&$24$&$48$&$ $&$ $&$ $&$ $&$ $ \\
			$n=5$&$8$&$48$&$192$&$ $&$ $&$ $&$ $ \\
			$n=6$&$48$&$0$&$1152$&$ $&$ $&$ $&$ $ \\
			$n=7$&$16$&$64$&$256$&$2816$&$ $&$ $&$ $ \\
			$n=8$&$96$&$-192$&$2688$&$768$&$ $&$ $&$ $ \\
			$n=9$&$32$&$64$&$384$&$4864$&$11264$&$ $&$ $ \\
			$n=10$&$192$&$-768$&$6912$&$-12288$&$141312$&$ $&$ $ \\
			$n=11$&$64$&$0$&$768$&$8192$&$6144$&$294912$&$ $ \\
			$n=12$&$384$&$-2304$&$18432$&$-61400$&$405504$&$-663552$&$ $ \\
			$n=13$&$128$&$-256$&$2048$&$12288$&$-12288$&$614400$&$-98304$ \\
			\hline \hline
		\end{tabular}
	\end{center}
\end{table}

Besides the property of the projector $P_{\lfloor n/2 \rfloor+i}^2$, there is also another important property of AME states,
\begin{eqnarray}
\rho_{(\lfloor n/2 \rfloor+i)}\otimes I^{\otimes (n-k)}|\psi\rangle=d^{-(n-k)}|\psi\rangle.\label{e1}
\end{eqnarray}
Recall that the density matrix of an AME state always has the following form:
\begin{eqnarray}
\begin{split}
&\rho_{(\lfloor n/2 \rfloor+i)}=d^{-(\lfloor n/2 \rfloor+i)}\\
&\left( I + \sum_{j_1\cdots j_{\lfloor n/2 \rfloor+1}}P_{\lfloor n/2 \rfloor+1}^{({j_1\cdots j_{\lfloor n/2 \rfloor+1}})}\otimes I^{\overline{(j_1\cdots j_{\lfloor n/2 \rfloor+1})}} + \sum_{j_1\cdots j_{\lfloor n/2 \rfloor+2}}P_{\lfloor n/2 \rfloor+2}^{({j_1\cdots j_{\lfloor n/2 \rfloor+2}})}\otimes I^{\overline{(j_1\cdots j_{\lfloor n/2 \rfloor+2})}} + \cdots +P_{(\lfloor n/2 \rfloor+i)}  \right)\label{e2}
\end{split}
\end{eqnarray}
(i.e. all operators $P_j$ with $1\le j \le \lfloor n/2 \rfloor$ vanish).

Inserting (\ref{e1}) into (\ref{e2}), for any $l=1,\cdots,i$, we have the following equation,
\begin{eqnarray}
\begin{split}
&d^{-(\lfloor n/2 \rfloor+l)}\left( \sum_{j_1\cdots j_{\lfloor n/2 \rfloor+1}}P_{\lfloor n/2 \rfloor+1}^{({j_1\cdots j_{\lfloor n/2 \rfloor+1}})}\otimes I^{\overline{(j_1\cdots j_{\lfloor n/2 \rfloor+1})}} + \cdots + P_{(\lfloor n/2 \rfloor+l)} \right)\otimes I^{\otimes (n-(\lfloor n/2 \rfloor+l))}|\psi\rangle\\
&=\left( d^{-(n-(\lfloor n/2 \rfloor+l))}-d^{-(\lfloor n/2 \rfloor+l)}\right)|\psi\rangle
.\label{e3}
\end{split}
\end{eqnarray}
One can see that $|\psi\rangle$ is the eigenvector of each $P_{(\lfloor n/2 \rfloor+l)}\otimes I^{\otimes (n-(\lfloor n/2 \rfloor+l))}$.
Suppose $P_{(\lfloor n/2 \rfloor+l)}\otimes I^{\otimes (n-(\lfloor n/2 \rfloor+l))}|\psi\rangle=\lambda_{(\lfloor n/2 \rfloor+l)}|\psi\rangle$.
Then equations (\ref{e3}) leads to the following linear equations,
\begin{equation}
 \left \{
 \arraycolsep=0pt
 \begin{array}{rrrrl}
 B_{11}x_1&&&=&R_1,\\
 B_{21}x_1&+B_{22}x_2&&=&R_2,\\
 \cdots&&&&\\
  B_{i1}x_1&+B_{i2}x_2&+\cdots&+B_{ii}x_i=&R_i\,\
  \end{array}
 \right.
 \end{equation}
 where $x_j=\lambda_{(\lfloor n/2 \rfloor+j)}$, $R_l= d^{-(n-(\lfloor n/2 \rfloor+l))}-d^{-(\lfloor n/2 \rfloor+l)}$, $j,l=1,\cdots,i$, and
 \begin{equation}
B_{lj}=\left\{
\begin{array}{ll}
d^{-\lfloor n/2 \rfloor-l}\binom{\lfloor n/2 \rfloor+l}{\lfloor n/2 \rfloor+j},~~~~~& 1\leq j\leq l\leq i,\\
0,&1\leq l<j\leq i.\\
\end{array}
\right.
\end{equation}

Let $B$ denote the coefficient matrix with elements $B_{lj}$. The inverse matrix $B^{-1}$ is given by
\begin{equation}
B_{lj}^{-1}=\left\{
\begin{array}{ll}
(-1)^{l+j}d^{\lfloor n/2 \rfloor+j}\binom{\lfloor n/2 \rfloor+l}{\lfloor n/2 \rfloor+j},~~~~&1\leq j\leq l\leq i,\\
0,&1\leq l<j\leq i.
\end{array}
\right.
\end{equation}
One can easily verify that for $l<j$, $(BB^{-1})_{lj}=0$, and for $l\geq j$,
\begin{eqnarray}
\begin{split}
(BB^{-1})_{lj}&=\sum_{k=j}^{l} B_{lk}B_{kj}^{-1}\\
&=d^{j-l}\sum_{k=j}^{l}(-1)^{k+j}\binom{\lfloor n/2 \rfloor+l}{\lfloor n/2 \rfloor+k}\binom{\lfloor n/2 \rfloor+k}{\lfloor n/2 \rfloor+j}\\
&=d^{j-l}\binom{\lfloor n/2 \rfloor+l}{\lfloor n/2 \rfloor+j}\sum_{k=j}^{l}(-1)^{k-j}\binom{l-j}{k-j}\\
&=d^{j-l}\binom{\lfloor n/2 \rfloor+l}{\lfloor n/2 \rfloor+j}(1-1)^{l-j}\\
&=\delta_{lj}.
\end{split}
\end{eqnarray}
And the constant vector is given as $R_l=d^{-(n-(\lfloor n/2 \rfloor+l))}-d^{-(\lfloor n/2 \rfloor+l)}$.(???)
Thus we have the following theorem

\begin{thm}\label{tracesquareSs}
Let $|\Phi\rangle$ be an AME state given in the form of (\ref{kuniformstte}). The eigenvalues of $\lambda_{(\lfloor n/2 \rfloor+i)}$ in
$P_{(\lfloor n/2 \rfloor+i)}\otimes I^{\otimes (n-(\lfloor n/2 \rfloor+i))}|\psi\rangle=\lambda_{(\lfloor n/2 \rfloor+i)}|\psi\rangle$ are given by
\begin{eqnarray}
\begin{split}
\lambda_{(\lfloor n/2 \rfloor+i)}&=\sum_{j=1}^{i}(B^{-1})_{ij}R_j \\
&=\frac{(-1)^i \binom{i+\lfloor n/2 \rfloor}{1+\lfloor n/2 \rfloor}(1+\lfloor n/2 \rfloor-d^{2(1+\lfloor n/2 \rfloor)-n}(i+\lfloor n/2 \rfloor){}_{2}F_{1}(1,1-i;2+\lfloor n/2 \rfloor;d^2))}{i+\lfloor n/2 \rfloor},\label{e4}
\end{split}
\end{eqnarray}
where ${}_{2}F_{1}(a,b;c;z)$ is the ordinary hyper-geometric function.
\end{thm}
\par

Notice that the only difference between the right hand sides of equations (\ref{p1}) and (\ref{e4}) is
the factor $d^{(\lfloor n/2 \rfloor+i)}$, which means that $\lambda_{(\lfloor n/2 \rfloor+i)}\geq 0$ for any AME states.


In summary, from the generalized Bloch representation of AME states,
we have proved that the trace of the squared support for any given weight
is determined and irrelevant with the choice of the subsystems, that is,
$tr(P_{\lfloor n/2\rfloor+i}^2)$ is an invariant for any $\lfloor n/2\rfloor+i$ subsystems.
Based on this fact, we have obtained an explicit formula on $tr(P_{\lfloor n/2\rfloor+i}^2)$ and $\lambda_{(\lfloor n/2 \rfloor+i)}$ for arbitrary AME states
given by the so-called hyper-geometric function. A deep connection between these two quantities is also obtained. Moreover, we find that $i=2$ is always optimal on
verifying the existence of AME states. That is, the results obtained in \cite{Huber2017} are already the best, where the authors only studied the case of $i=2$.
Further more, it has been also shown that the eigenvalues of the projectors are always positive for any AME states. Our results
improve the knowledge on the non-existence of AME states for given $d$ and $n$, and may be used to provide improved criterion on the existence of AME
states, as well as benefit to the construction of the so-called $k$-uniform ($k<\lfloor n/2\rfloor$) states \cite{Goyeneche2014}.

\bigskip
\noindent {\bf Acknowledgments}
We thank Felix Huber for helpful discussions.  This work was completed while Bo Li was visiting the
Max-Planck-Institute for Mathematics in the Sciences in Germany under the support of the China Scholarship
Council (Grant No. 201608360191). This work is supported by NSFC(11765016,11675113) and Jiangxi Education Department Fund (KJLD14088).


\begin{thebibliography}{18}
\bibitem{Nielsen&Chuang} M. A. Nielsen and I.~L. Chuang, \emph{Quantum Computation
and Quantum Information} (Cambridge University Press,
Cambridge, England, 2000).

\bibitem{Gisin1998}
  N.\ Gisin and H.\ Bechmann-Pasquinucci,
  Phys. Lett. A {\bf 246}, 1  (1998).

\bibitem{Higuchi2000}
  A.\ Higuchi and A.\ Sudbery,
  Phys. Lett. A {\bf 273}, 213 (2000).

  \bibitem{Helwig2012}
  W. Helwig, W. Cui, J. I. Latorre, A. Riera, and H.-K. Lo,
  Phys. Rev. A {\bf 86}, 052335 (2012).

  \bibitem{Helwig2013}
  W. Helwig and W. Cui,
  arXiv:1306.2536.

\bibitem{Helwig2013_graph}
 W. Helwig,
 arXiv:1306.2879.

 \bibitem{Hein2006}
  M. Hein, J. Eisert, and H.J. Briegel,
  Phys. Rev. A {\bf 69}, 062311 (2004).

\bibitem{Steinhoff}
F. E.S. Steinhoff, C. Ritz, N. Miklin, O. G¨¹hne
Phys. Rev. A {\bf 95}, 052340 (2017)



\bibitem{Chen2016}
  L. Chen and D. L. Zhou,
  Scientific Reports {\bf 6}, 27135 (2016).



\bibitem{Scott2004}
  A.\ J.\ Scott,
  Phys. Rev. A {\bf 69}, 052330 (2004).

  \bibitem{raissi}
  Z. Raissi, C. Gogolin, A. Riera, A. Ac\'{\i}n,
 arXiv:1701.03359.


\bibitem{Nebe2006}
  G. Nebe, E. M. Rains, and N. J. A. Sloane,
  {\it Self-Dual Codes and Invariant Theory},
  Springer Berlin-Heidelberg (2006).



\bibitem{Grassl2015}
  M.\ Grassl and M.\ Roetteler,
  Proceedings 2015 IEEE International Symposium on Information Theory, {\bf14-19} 1104(2015)

\bibitem{Calderbank1998}
  A.\ R.\ Calderbank, E.\ M.\ Rains, P.\ W.\ Shor, and N.\ J.\ A.\ Sloane,
  IEEE Trans. Inf. Theory {\bf 44}, 1369 (1998)

  \bibitem{Zha2011}
  X.-W. Zha, H.-Y. Song, J.-X. Qi, D. W., and Q. Lan,
  J. Phys. A: Math. Theor. {\bf 45}, 255302 (2012);
   X. Zha, C. Yuan and Y. Zhang, Laser Phys. Lett. {\bf 10}, 045201  (2013).


 \bibitem{Goyeneche2015}
  D. Goyeneche, D. Alsina, J. I. Latorre, A. Riera, and K. {\.Z}yczkowski,
  Phys. Rev. A {\bf 92}, 032316 (2015).

 \bibitem{Latorre2015}
  J. I. Latorre and G. Sierra, 
arXiv:1502.06618 (2015).

 \bibitem{Pastawski}
  F. Pastawski, B. Yoshida, D. Harlow and J.
arXiv:1503.06237 (2015).

\bibitem{Arnoud2013}
  L. Arnaud and N. J. Cerf,
  Phys. Rev. A {\bf 87}, 012319 (2013).

\bibitem{Goyeneche2014}
  D.\ Goyeneche and K.\ {\.Z}yczkowski,
  Phys.\ Rev.\ A {\bf 90}, 022316 (2014).

  \bibitem{Feng2015}
  K. Feng, L. Jin, C. Xing, and C. Yuan,
  arXiv:1511.07992.

  \bibitem{Bernal2017}
  A. Bernal,
  Quant. Phys. Lett. {\bf 6}, 1 (2017)

  \bibitem{Gour2010}
  G. Gour and N. R. Wallach,
  J. Math. Phys. {\bf 51}, 112201 (2010).


\bibitem{Borras2007}
  A. Borras, A. R. Plastino, J. Batle, C. Zander, M. Casas, and A. Plastino,
  J. Phys. A: Math. Theor. {\bf 44}, 13407 (2007).

  \bibitem{Facchi2010}
  P. Facchi, G. Florio, U. Marzolino, S. Pascazio, and G. Parisi,
  J.\ Phys.\ A: Math.\ Theor.\ {\bf 43},  225303  (2010); P. Facchi, G. Florio, G. Parisi, and S. Pascazio,
   Phys. Rev. A {\bf 77}, 060304 (2008).

  \bibitem{Rains1998}
  E. M. Rains,
  IEEE Trans. Inf. Theory {\bf 44}, 1388 (1998).

\bibitem{Rains1999}
  E. M. Rains,
  IEEE Trans. Inf. Theory {\bf 45}, 2361 (1999).

  \bibitem{Huber2017}
  F. Huber, O. G\"{u}hne, and J. Siewert,
  Phys. Rev. Lett {\bf 118}, 200502 (2017).

  \bibitem{Hypergeometric} See the term ``Hypergeometric function" in Wikipedia.






























\end{thebibliography}
\end{document}